\documentclass[12pt,preprint]{aastex}

\shorttitle{\indent \def Magnetic reconnection at flux rope
boundaries} \shortauthors{Tian et al.}

\begin{document}

\title{Signatures of magnetic reconnection at boundaries of interplanetary small-scale magnetic flux ropes}

\author{Hui Tian\altaffilmark{1}, Shuo Yao\altaffilmark{1}, Qiugang Zong\altaffilmark{1}, Jiansen He\altaffilmark{2}, Yu Qi\altaffilmark{1}}

\altaffiltext{1}{School of Earth and Space Sciences, Peking
University, 100871 Beijing, China; tianhui924@pku.edu.cn}

\altaffiltext{2}{Max-Planck-Institut f\"ur Sonnensystemforschung,
37191 Katlenburg-Lindau, Germany}

\begin{abstract}
The interaction between interplanetary small-scale magnetic flux
ropes and the magnetic field in the ambient solar wind is an
important topic to understanding the evolution of magnetic
structures in the heliosphere. Through a survey of 125 previously
reported small flux ropes from 1995 to 2005, we find that 44 of them
reveal clear signatures of Alfv\'{e}nic fluctuations, and thus
classify them into Alfv\'{e}n wave trains rather than flux ropes.
Signatures of magnetic reconnection, generally including a plasma
jet of $\sim$30~km~s$^{-1}$ within a magnetic field rotational
region, are clearly present at boundaries of about 42\% of the flux
ropes and 14\% of the wave trains. The reconnection exhausts are
often observed to show a local increase in the proton temperature,
density and plasma beta. About 66\% of the reconnection events at
flux rope boundaries are associated with a magnetic field shear
angle larger than 90$^\circ$ and 73\% of them reveal a decrease by
20\% or more in the magnetic field magnitude, suggesting a dominance
of anti-parallel reconnection at flux rope boundaries. The
occurrence rate of magnetic reconnection at flux rope boundaries
through the year of 1995 to 2005 is also investigated and we find
that it is relatively low around solar maximum and much higher when
approaching solar minima. The average magnetic field depression and
shear angle for reconnection events at flux rope boundaries also
reveal a similar trend from 1995 to 2005. Our results demonstrate
for the first time that boundaries of a substantial fraction of
small-scale flux ropes have properties similar to those of magnetic
clouds, in the sense that both of them exhibit signatures of
magnetic reconnection. The observed reconnection signatures could be
related either to the formation of small flux ropes, or to the
interaction between flux ropes and the interplanetary magnetic
fields.

\end{abstract}


\section{Introduction}

Magnetic flux ropes, which are usually associated with eruptive
processes in space and astrophysical plasmas, are helical magnetic
field structures
\citep[e.g.,][]{Shitaba1986,Zong2004,Demoulin2009,Linton2009,Cheng2010}.
Interplanetary magnetic flux ropes can be classified into two
categories, the large-scale ($\sim$ day) magnetic clouds and
small-scale ($\sim$ hour) flux ropes \citep{Cartwright2008}.
Magnetic clouds, which are characterized by their high magnetic
field magnitude, low proton temperature, and smooth rotation of the
magnetic field direction through a large angle on the time scale of
one day
\citep[e.g.,][]{Burlaga1981,Schwenn1991,Lepping2006,Jian2006}, are
believed to be a subset of interplanetary coronal mass ejections
(ICMEs). The small-scale flux ropes, which exhibit a similar
magnetic field behavior to their larger-scale counterparts, were
discovered by \cite{Moldwin1995} and \cite{Moldwin2000} through
observations of the Ulysses, IMP 8, and Wind spacecraft. Unlike
magnetic clouds, the proton density is often not depressed inside
these small flux ropes. The small flux ropes are more likely to be
observed in the slow rather than fast solar wind
\citep{Cartwright2008,Feng2008}. The small flux ropes are suggested
to be the interplanetary manifestations of small-scale solar
eruptions \citep{Tu1997,Mandrini2005,Feng2007,Feng2008,Wu2008} or
the products of local magnetic reconnection in the solar wind
\citep{Moldwin2000,Cartwright2008,Ruan2009}.

Magnetic reconnection is an important mechanism of energy conversion
in space and astrophysical plasmas
\citep[e.g.,][]{Shitaba1994,Priest2000,Deng2004,Xiao2006,He2008,He2010}.
Although it has been suggested or predicted by many authors that
such a process should be prevalent in the solar wind since tens of
years ago, the observational signatures of magnetic reconnection has
not been identified and recognized until recently.
\cite{Gosling2005a} identified Petschek-type reconnection exhausts
which are characterized by a roughly Alfv\'{e}nic accelerated plasma
jet confined to a region with rotational magnetic field and bounded
by correlated changes in components of the magnetic field \textbf{B}
and flow velocity \textbf{V} on one side and anti-correlated changes
on the other. In the reconnection exhaust, the magnetic field
magnitude is depressed, and as additional signatures the proton
density, temperature and plasma beta are often locally enhanced.
Subsequent intensive studies suggest that reconnection in the solar
wind is often associated with extended X-lines
\citep{Phan2006,Gosling2007b} and are predominantly observed in the
low-speed rather than high-speed solar wind
\citep{Gosling2006b,Gosling2007d,Phan2009}. Only a few cases of
reconnection have been found to be associated with the heliospheric
current sheet \citep{Gosling2005b,Gosling2006a,Gosling2007c}. Local
reconnection in the solar wind is found to be not efficient to
produce energetic particles \citep{Gosling2005c}. Reconnection is
found to be prevalent at small shear angles of the magnetic field
\citep{Gosling2007a,Gosling2008} in the low-beta solar wind
\citep{Phan2009}.

Numerical simulations have shown that reconnection between CME flux
ropes and the ambient magnetic field may play an important role in
the evolution of CMEs \citep[e.g.,][]{Shiota2010}.
\cite{McComas1994} suggested that magnetic reconnection should
commonly occur at the interfaces between fast ICMEs and the ambient
solar wind. Some reported reconnection exhausts were indeed
associated with ICMEs \citep[e.g.,][]{Gosling2005a}. \cite{Wei2003}
found that most magnetic clouds have boundary layers displaying a
drop in the magnetic field magnitude and a significant change of the
field direction, as well as properties of a high proton temperature,
density and plasma beta. They claimed that these signatures are
manifestations of magnetic reconnection through interaction between
magnetic clouds and the ambient medium.

The interaction between interplanetary small-scale magnetic flux
ropes and the magnetic field in the ambient solar wind is also an
important topic but remains poorly understood. \cite{Feng2009}
reported a small flux rope followed by a reconnection exhaust.
However, until now no systematic study of the boundaries of small
flux ropes has been done. In this paper we perform the first
statistical investigation of the boundaries of interplanetary
small-scale flux ropes, and identify signatures of magnetic
reconnection between the flux ropes and the interplanetary magnetic
field.

\section{Case studies and discussion}

\cite{Feng2008} performed a visual survey of the 1-minute averaged
magnetic field and plasma data obtained by the Wind satellite and
identified 125 small interplanetary magnetic flux ropes from 1995 to
2005 with the help of the Lundquist fitting technique
\citep{Lundquist1950,Goldstein1983}. We started with this list of
small flux ropes and investigated properties of the rope boundaries
by using the 3-s magnetic field and plasma data obtained by the MFI
\citep{Lepping1995} and 3DP \citep{Lin1995} instruments onboard
Wind. A reconnection event (exhaust) was identified using the
following primary criteria: (1) an obvious plasmas jet within a
region with rotational magnetic field, (2) at least one component of
both the velocity change and magnetic field change are correlated on
one side and anti-correlated on the other side. As an interesting
result, we found that clear signatures of magnetic reconnection are
present at one or both boundaries of about 42\% of the small flux
ropes. Below we show several examples before presenting the
statistical results.

\subsection{Magnetic reconnection at both boundaries of the flux rope}

\begin{figure}
\centering {\includegraphics[width=\textwidth]{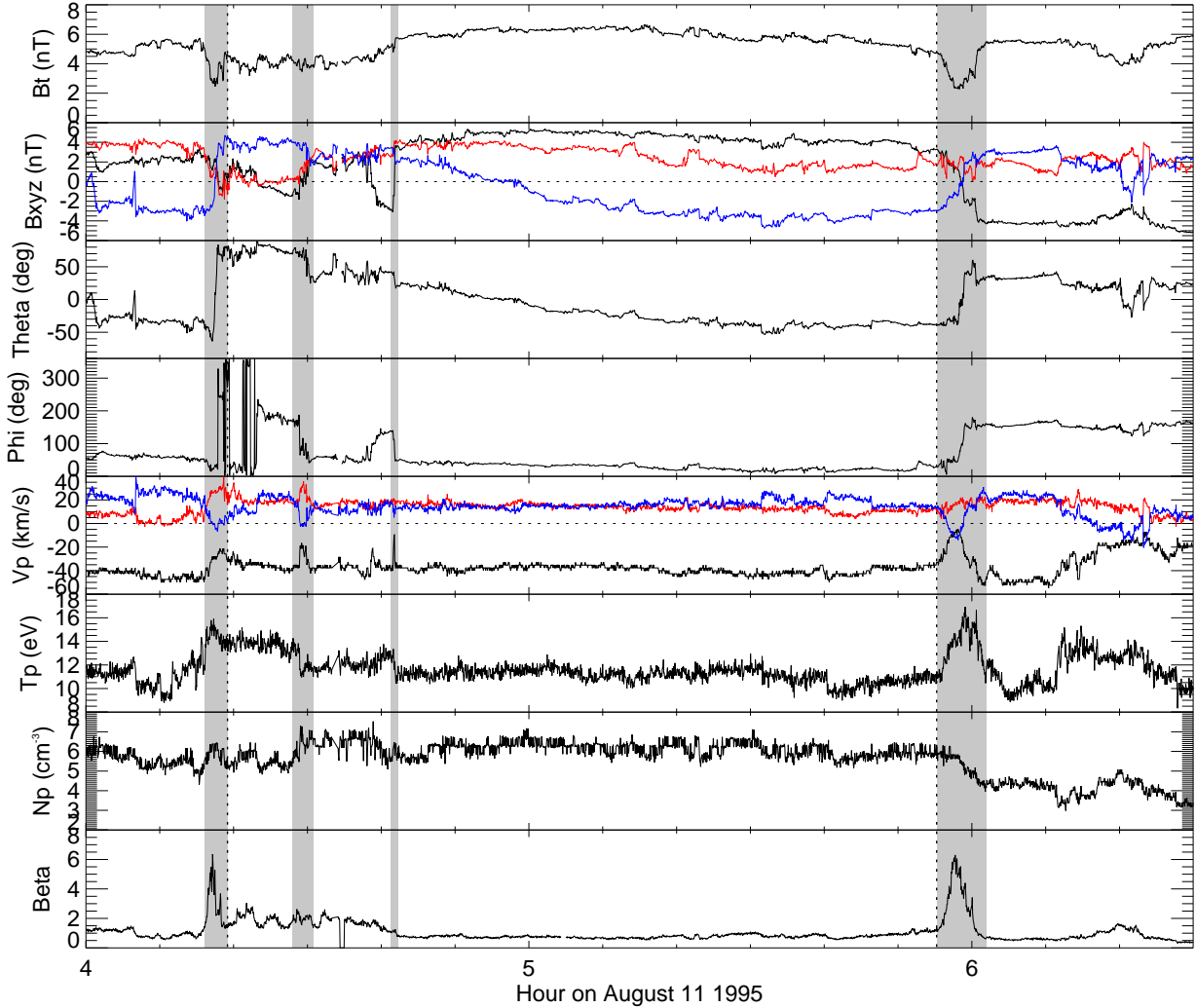}}
\caption{Magnetic field and plasma parameters for an interplanetary
small-scale magnetic flux rope observed by Wind on August 11 1995.
From top to bottom: magnetic field magnitude (Bt), magnetic field
vector in GSE cartesian coordinates (Bxyz), elevation angle of the
magnetic field (Theta), azimuthal angle of the magnetic field (Phi),
proton velocity vector in GSE cartesian coordinates (Vxyz), proton
temperature (Tp), proton number density (Np), and plasma beta
(Beta). The x, y and z-components of the magnetic field and proton
velocity are denoted by the black, red and blue lines, respectively.
A value of 450~km~s$^{-1}$ has been added to the x-component of the
velocity. The two dotted vertical lines mark the beginning and end
of the flux rope. The shaded regions correspond to Wind crossing of
reconnection exhausts. } \label{fig.1}
\end{figure}

\begin{figure}
\centering {\includegraphics[width=0.48\textwidth]{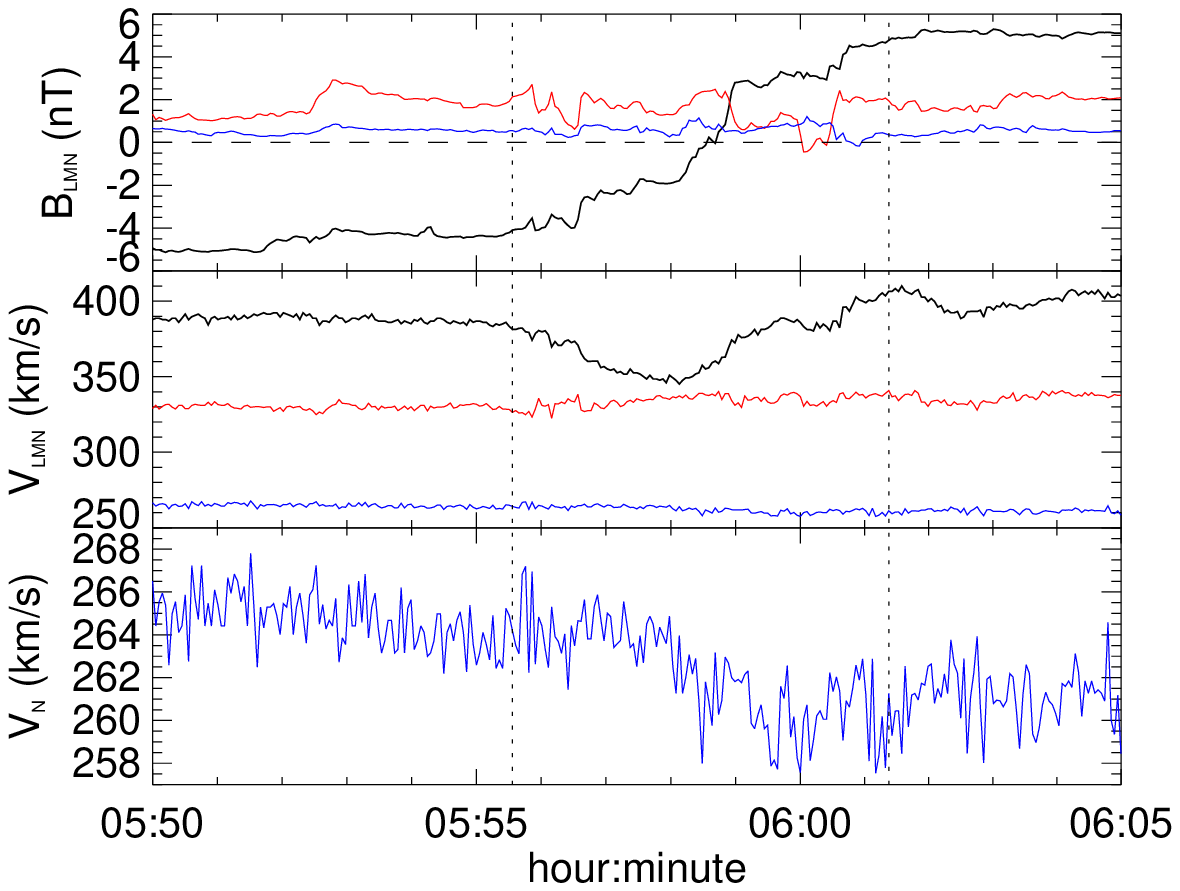}}
\caption{A reconnection exhaust observed by Wind around 06:00 on
August 11 1995. From top to bottom: magnetic field vector
(B$_{\rm{LMN}}$), velocity vector (V$_{\rm{LMN}}$) and the
N-component of the velocity (V$_{\rm{N}}$). The L, M, and
N-components of the magnetic field and proton velocity are denoted
by the black, red and blue solid lines, respectively. The two dotted
vertical lines mark the beginning and end of the reconnection
exhaust. A value of 200~km~s$^{-1}$ has been added to the
M-component of the velocity. } \label{fig.2}
\end{figure}

Figure~\ref{fig.1} shows the magnetic field and plasma parameters
for a typical small-scale flux rope observed on August 11, 1995. The
two dotted vertical lines mark the beginning and end of the flux
rope. Primary signatures of reconnection exhausts, plasma jets
within the magnetic field rotational region, were clearly present at
both boundaries of this flux rope.

The exhaust at the trailing boundary was observed from about 05:55
to 06:02. A strong decrease of the magnetic field magnitude (57\%)
was observed inside the exhaust. The plasma jets and magnetic field
rotation mainly occurred in the x and z-directions in the GSE
coordinate system. In these two directions, we can see a clear
anti-correlation between the changes of \textbf{B} and \textbf{V} at
the front side and a correlation at the tail side of the exhaust.
The shear angle of the magnetic field across the exhaust was
relatively large (133$^\circ$). Additional signatures of magnetic
reconnection, including a local enhancement of the proton
temperature and plasma beta, were also observed in the exhaust. The
proton density was not enhanced but intermediate to those on the
opposite sides of the exhaust, indicating that the transitions from
the ambient medium to the exhaust are not slow-mode-like on both
sides \citep{Gosling2006b}.

We performed a minimum variance analysis (MVA) of the magnetic field
from 05:50 to 06:05, and established the LMN coordinate system for
the reconnecting current sheet
\citep{Sonnerup1967,Davis2006,Phan2006}. The magnetic field rotated
predominantly in the L direction, which was [-0.78,0.01,0.63] in GSE
cartesian coordinates. The M direction aligning with the X-line
orientation was found to be [-0.26,0.91,-0.32] in GSE. The overall
current sheet normal, the N direction, was [0.57,0.41,0.71] in GSE.
Figure~\ref{fig.2} shows the magnetic fields and proton velocities
as converted to this LMN coordinates. In the normal direction to the
current sheet, the proton velocity across the exhaust is
differentiated by about 4~km~s$^{-1}$, indicating an inflow velocity
of 2~km~s$^{-1}$ in the frame of the current sheet. The reconnection
electric field was estimated to be 0.01~mV~m$^{-1}$, with a magnetic
field of about 5~nT convecting into the reconnection region at this
inflow speed. The dimensionless reconnection rate, the inflow speed
divided by the Alfv\'{e}n speed, was calculated to be about 5\%,
indicating fast reconnection \citep{Davis2006,Phan2006}.

In the list of \cite{Feng2008}, the duration of this flux rope was
from 04:41 to 05:56. From Figure~\ref{fig.1} we can clearly see a
reconnection exhaust just at about 04:41. This exhaust only lasted
for about 18~s, and the magnetic field reversal and plasma jet were
mainly present in the x-direction. However, if we look at the
overall behavior of the magnetic field elevation angle of this flux
rope, we can see an almost smooth variation starting at 04:19. From
about 04:16 to 04:19, an abrupt large rotation of the elevation
angle accompanied by all typical signatures of magnetic reconnection
were observed. A strong depression of the magnetic field magnitude
(44\%) and a large shear angle (143$^\circ$) of the magnetic field
vectors were associated with this reconnection exhaust. We think
that it is more appropriate to use this reconnection event as the
leading boundary of the flux rope. In fact, the leading boundary of
this flux rope seems to be a complex layer rather than a simple
edge. This layer may correspond to the time interval from 04:16 to
04:41. Between the reconnection exhaust around 04:19 and the one
around 04:41, we even found a third possible reconnection exhaust at
about 04:29 within the leading boundary layer. These findings
suggest that the interaction between small-scale flux ropes and the
ambient solar wind can be very complicated.

\begin{figure}
\centering {\includegraphics[width=\textwidth]{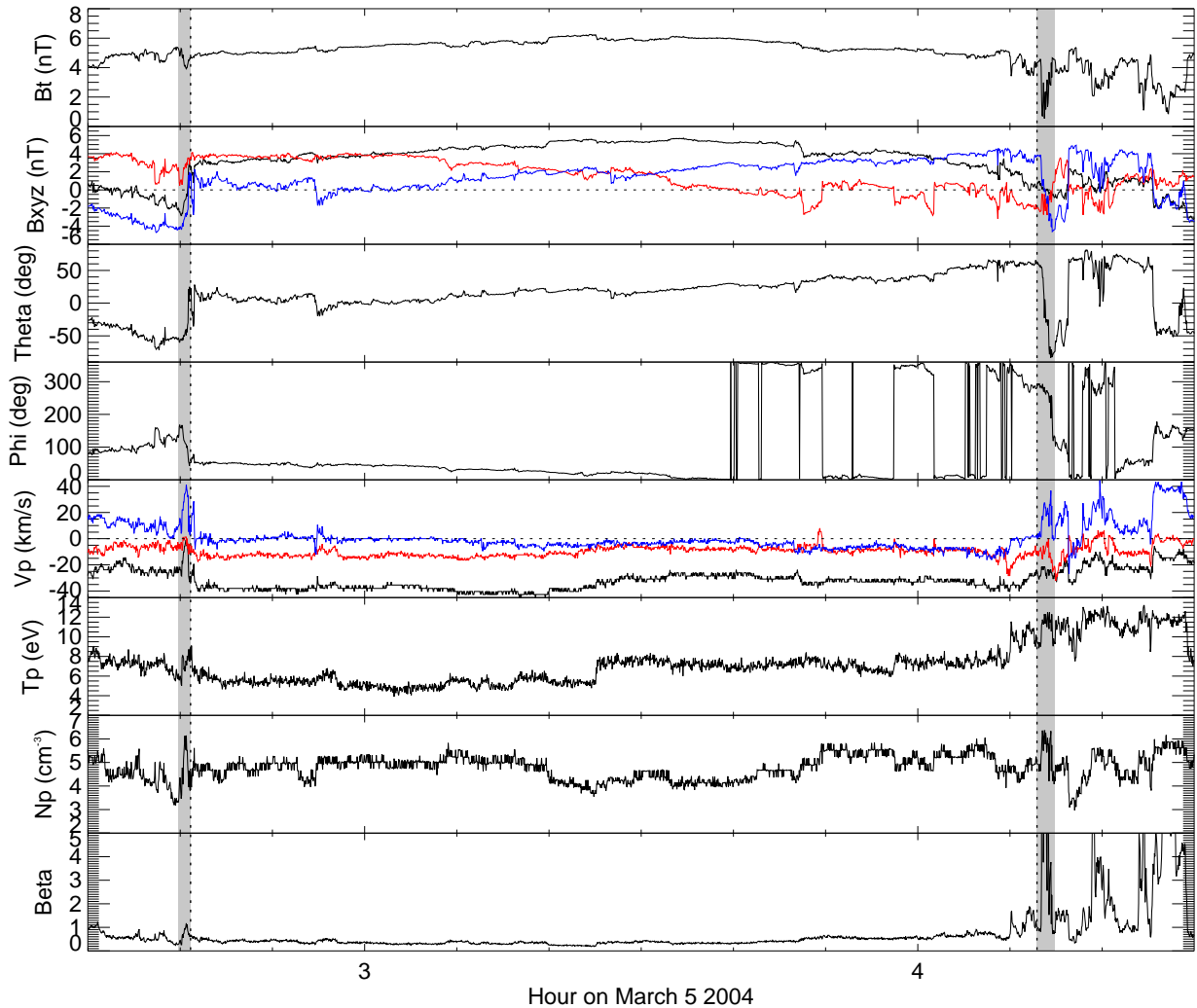}}
\caption{Magnetic field and plasma parameters for an interplanetary
small-scale magnetic flux rope observed by Wind on March 5 2004. A
value of 450~km~s$^{-1}$ has been added to the x-component of the
velocity. } \label{fig.3}
\end{figure}

Another example of magnetic reconnection at both boundaries of a
flux rope is presented in Figure~\ref{fig.3}. This flux rope was
encountered by Wind on March 5, 2004. According to \cite{Feng2008},
the duration of this flux rope was from 02:41 to 04:12. A
reconnection exhaust occurring at about 02:41 was present just ahead
of the flux rope, while another one starting at about 04:13 was
observed just behind the flux rope. At the leading boundary the jet
was found in the x and z-directions in GSE coordinate system, while
at the trailing boundary the jet was mainly found in the
z-direction. A relatively weak (26\%) depression of the magnetic
field magnitude was present in the leading boundary, while at the
trailing boundary a very strong (75\%) decrease was observed. The
magnetic field shear angles across the two exhausts were found to be
92$^\circ$ and 161$^\circ$, respectively. Both exhausts revealed a
local enhancement in the proton temperature, density, and plasma
beta, consistent with Petschek's theory that the reconnection
exhausts are bounded by slow-mode wave on both sides
\citep{Petschek1964}.

\subsection{Magnetic reconnection at one boundary of the flux rope}

Signatures of magnetic reconnection were not always present on both
boundaries of small-scale flux ropes. In many cases, reconnection
exhausts were only observed to be associated with the leading or
trailing boundaries of flux ropes.

\cite{Feng2009} reported a small flux rope followed by a
reconnection exhaust, and suggested that the magnetic fields of the
flux rope were peeled off through the reconnection between the flux
rope and the background solar wind. However, it seems that this
event was inside the leading boundary layer of an near-Earth ICME
starting at about 13:00 on March 25 and ending at 10:00 on March 26
in 1998 \citep{Cane2003}. This small flux rope might correspond to
the second flux rope produced along with the CME flux rope as
predicted by the breakout model of \cite{vanderHolst2007}. And the
reconnection might be the result of the interaction between the
small flux rope and the ICME.

\begin{figure}
\centering {\includegraphics[width=\textwidth]{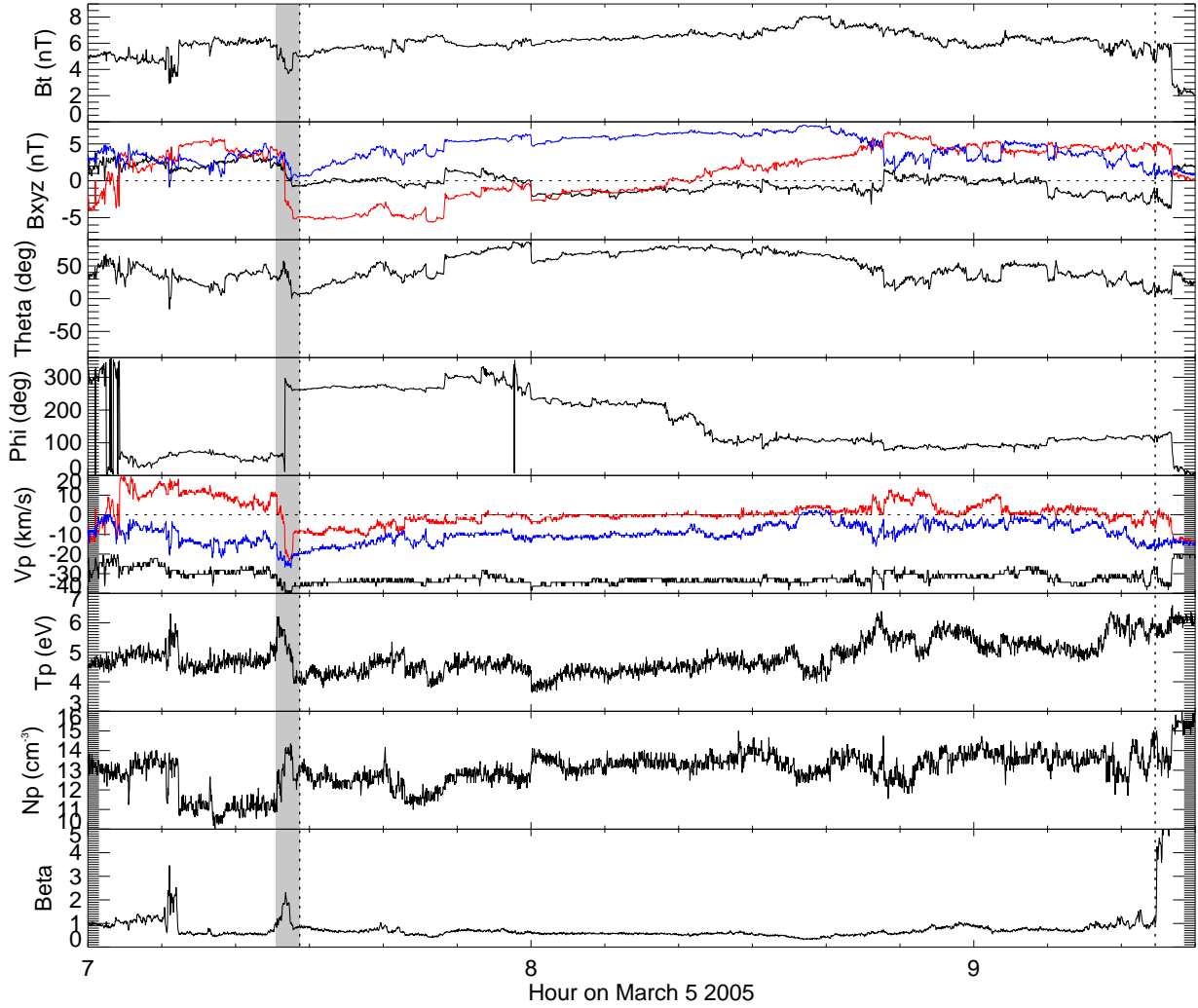}}
\caption{Magnetic field and plasma parameters for an interplanetary
small-scale magnetic flux rope observed by Wind on March 5 2005. A
value of 380~km~s$^{-1}$ has been added to the x-component of the
velocity. } \label{fig.4}
\end{figure}

A reconnection exhaust at the leading boundary of a flux rope
observed by Wind on March 5 2005 is shown in Figure~\ref{fig.4}. The
reconnection exhaust was encountered by Wind within 80~s, beginning
at 07:26:35. A rotation of the magnetic field direction can be
identified mainly from the y-component. Correspondingly, an obvious
jetting plasma was present in the y-direction. The changes of the
magnetic field and proton velocity were correlated on the front side
and anti-correlated on the tail side of the exhaust. Obvious
enhancements of the proton density, temperature, and plasma beta
were all observed inside the exhaust. The magnetic field magnitude
was decreased by 33\% in the exhaust. After crossing the exhaust,
the magnetic field direction changed by 135$^\circ$. No reconnection
exhaust was observed at the trailing boundary of this flux rope.

\begin{figure}
\centering {\includegraphics[width=\textwidth]{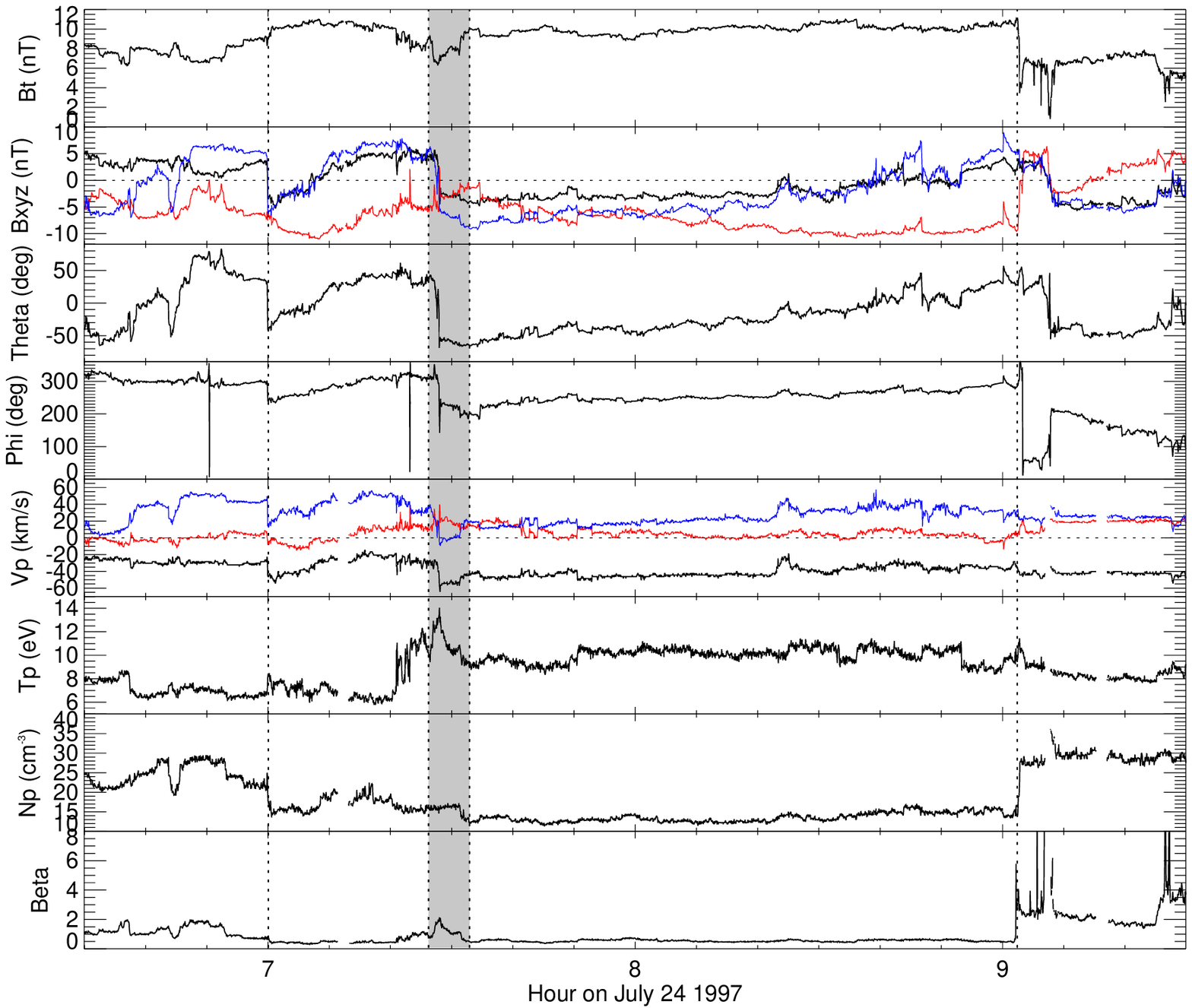}}
\caption{Magnetic field and plasma parameters for two interplanetary
small-scale magnetic flux ropes observed by Wind on July 24 1997. A
value of 370~km~s$^{-1}$ has been added to the x-component of the
velocity. A reconnection exhaust was observed between the two flux
ropes.} \label{fig.5}
\end{figure}

In the list of \cite{Feng2008}, there are two flux ropes which were
observed by Wind one after the other on July 24, 1997.
Figure~\ref{fig.5} shows the magnetic field and plasma parameters of
these two cases. The first flux rope encountered by Wind from about
07:00 to 07:26, and the second one from 07:32 to 9:02.
Interestingly, we found clearly a reconnection exhaust between these
two flux ropes. The plasma jet was primarily observed in the x and
z-components. The magnetic field decreased by about 22\% inside the
exhaust. And a shear angle of 132$^\circ$ was observed across the
exhaust. The exhaust was immediately preceded by the first rope, and
immediately followed by the second one. The bipolar signatures of
the two flux ropes were both present in the x and z-directions,
indicating a similar orientation of the two flux ropes. Indeed, the
longitude and latitude angles of the two ropes' axes in the ecliptic
coordinate system, which were [209$^\circ$,-12$^\circ$] and
[281$^\circ$,-6$^\circ$] as obtained from the fitting with the
Lundquist solution in \cite{Feng2008}, were not so different.
Imagining two adjacent flux ropes with the same handedness aligned
in a similar orientation, we can find that an almost anti-parallel
magnetic field configuration will form between the two ropes. This
configuration is favorable for the occurrence of magnetic
reconnection. So the reconnection event we observed seems to be the
result of interactions between the two small-scale flux ropes.

\section{Statistical results and discussion}

We surveyed the entire list of 125 flux ropes in \cite{Feng2008},
and identified 47 reconnection exhausts at one or two boundaries of
these possible flux ropes. The starting and ending time of the
reconnection events and their associated flux ropes are listed in
Table~\ref{table1}. The decrease of the magnetic field magnitude
inside the reconnection exhausts (\%) and the shear angles of
magnetic field across the reconnection exhausts ($^\circ$) are also
shown there.

\begin{table}
\small 
\caption{Reconnection exhausts at boundaries of
interplanetary small-scale magnetic flux ropes.} \label{table1}
\centering
\begin{tabular}{cccccccccc}
\hline Number & Year & \textit{T$_{fb}$\footnotemark[1]} &
\textit{T$_{fe}$\footnotemark[2]} & \textit{L\footnotemark[3]} &
\textit{T$_{rb}$\footnotemark[4]} &
\textit{T$_{re}$\footnotemark[5]} & \textit{D\footnotemark[6]} &  \textit{S\footnotemark[7]} &  \textit{A\footnotemark[8]} \\
\hline
 1 &1995 &03072317 &03080043 &T &004330 &004830 &46 & 56 &A \\
 2 &1995 &03241125 &03241615 &L &112000 &112430 & 7 & 40 &F \\
 3 &1995 &05131025 &05131625 &L &102330 &102435 &45 & 54 &F \\
 4 &1995 &06172140 &06180441 &L &213850 &214010 &31 &155 &F \\
 5 &1995 &06172140 &06180441 &T &044025 &044105 &16 & 62 &F \\
 6 &1995 &08110419 &08110555 &L &041555 &041900 &44 &143 &F \\
 7 &1995 &08110419 &08110555 &T &055500 &060200 &57 &133 &F \\
 8 &1995 &08151419 &08151800 &L &141200 &141900 &50 &115 &F \\
 9 &1995 &08151419 &08151800 &T &180025 &180245 &29 &139 &F \\
10 &1995 &09201301 &09201401 &L &125730 &130030 &70 &156 &F \\
11 &1995 &09210254 &09210454 &L &025240 &025320 &40 &111 &F \\
12 &1996 &02101741 &02102207 &T &220730 &220845 & 4 & 56 &F \\
13 &1996 &03081947 &03090235 &T &023545 &023645 &11 &104 &F \\
14 &1996 &03130927 &03131021 &L &092540 &092630 &86 &169 &F \\
15 &1996 &05020541 &05020647 &L &054010 &054030 &50 &159 &F \\
16 &1996 &05170101 &05170955 &T &095500 &095545 &27 & 79 &F \\
17 &1996 &09281234 &09281440 &L &123145 &123400 &42 &121 &F \\
18 &1996 &09281234 &09281440 &T &144025 &144125 &71 &168 &F \\
19 &1997 &05111141 &05111401 &T &140145 &140310 &19 &123 &F \\
20 &1997 &05120524 &05120741 &T &074115 &074230 &20 & 51 &F \\
21 &1997 &05230616 &05231218 &L &061400 &061550 &54 & 68 &F \\
22 &1997 &05240222 &05240740 &T &073955 &074010 &21 & 71 &F \\
23 &1997 &05251910 &05260244 &L &190900 &190950 &29 & 98 &A \\
24 &1997 &07240700 &07240726 &T &072630 &073230 &22 &132 &F \\
25 &1997 &07240732 &07240902 &L &072630 &073230 &22 &132 &F \\
26 &1998 &03251328 &03251616 &T &161600 &162300 &41 &154 &F \\
27 &1998 &06021030 &06021635 &T &163505 &163610 &22 & 95 &A \\
28 &1998 &06260004 &06260750 &T &075000 &075015 &13 & 91 &F \\
29 &2000 &04181524 &04181743 &L &152325 &152345 & 7 & 43 &A \\
30 &2000 &04210635 &04210932 &L &063300 &063500 &13 & 84 &A \\
31 &2000 &08230923 &08231158 &T &115850 &115915 & 9 & 14 &F \\
32 &2000 &12231554 &12232137 &L &155420 &155520 & 9 & 85 &F \\
33 &2001 &01090242 &01090325 &T &032500 &032550 &13 &147 &F \\
34 &2001 &09261026 &09261717 &L &102500 &102530 & 4 & 18 &F \\
35 &2001 &09261026 &09261717 &T &171700 &171730 &14 & 89 &F \\
36 &2002 &01192108 &01192145 &T &214450 &214630 &32 &142 &F \\
37 &2002 &02182306 &02190249 &L &230135 &230535 & 8 & 93 &F \\
38 &2002 &03171932 &03172116 &L &193010 &193130 & 8 & 66 &F \\
39 &2003 &04291641 &04291852 &L &163930 &164050 &40 &120 &F \\
40 &2003 &04291641 &04291852 &T &185200 &190700 &40 & 93 &F \\
41 &2003 &08050824 &08051021 &L &081700 &082400 &20 &114 &A \\
42 &2003 &09292209 &09300134 &T &013450 &013650 &22 & 71 &F \\
43 &2004 &03050241 &03050413 &L &024005 &024100 &26 & 92 &F \\
44 &2004 &03050241 &03050413 &T &041315 &041450 &75 &161 &F \\
45 &2004 &08022335 &08030100 &L &233300 &233410 &21 &120 &F \\
46 &2005 &03050728 &03050924 &L &072635 &072755 &33 &135 &F \\
47 &2005 &11110011 &11110052 &T &005230 &010000 &70 &143 &F \\
\hline
\end{tabular}
\end{table}
\footnotetext[1]{The beginning of the flux ropes, MonthDayHourMinute
UT.} \footnotetext[2]{The end of the flux ropes, MonthDayHourMinute
UT.} \footnotetext[3]{The leading (L) or trailing (T) boundaries of
the flux ropes.} \footnotetext[4]{The beginning of the reconnection
exhausts, HourMinuteSecond UT.} \footnotetext[5]{The end of the
reconnection exhausts, HourMinuteSecond UT.} \footnotetext[6]{The
decrease of the magnetic field inside the reconnection exhausts
(\%).} \footnotetext[7]{The shear angles of magnetic field across
the reconnection exhausts ($^\circ$).} \footnotetext[8]{Identified
as flux ropes (F) or Alfv\'{e}n wave trains (A).}

\subsection{Detection of Alfv\'{e}nic fluctuations}

\cite{Cartwright2008} mentioned that some Alfv\'{e}n wave trains
might show observational magnetic field properties similar to flux
ropes. An inspection of the magnetic field and velocity data of flux
ropes listed in \cite{Feng2008} suggested that some of these flux
ropes revealed clear signatures of Alfv\'{e}n waves. We designed an
automatic algorithm to recognize these wave trains. We first
smoothed the magnetic field \textbf{B} and proton velocity
\textbf{V} over 20 minutes, respectively. Fluctuations of the
magnetic field and velocity were then obtained by subtracting the
smoothed data from the original data. The resulting magnetic field
perturbations were used to calculate the Alfv\'{e}n velocity
fluctuations using the formula
\textbf{$\delta$V}$_b$=\textbf{$\delta$B}/$\sqrt{\mu\rho}$. Here
\textbf{$\delta$B}, $\mu$ and $\rho$ refer to the magnetic field
perturbation, permeability of free space and proton mass density.
Finally, the correlation coefficients between components of the
Alfv\'{e}n velocity and proton velocity perturbations were
calculated. Flux ropes listed in \cite{Feng2008} were determined to
be Alfv\'{e}n wave trains if two or three of the correlation
coefficients are larger than 0.6. This method is similar to
\cite{Denskat1977} except for the inclusion of the density effect
here. Alfv\'{e}n wave trains and the rest flux ropes are marked in
Table~\ref{table1} as "A" and "F", respectively.

\begin{figure}
\centering {\includegraphics[width=\textwidth]{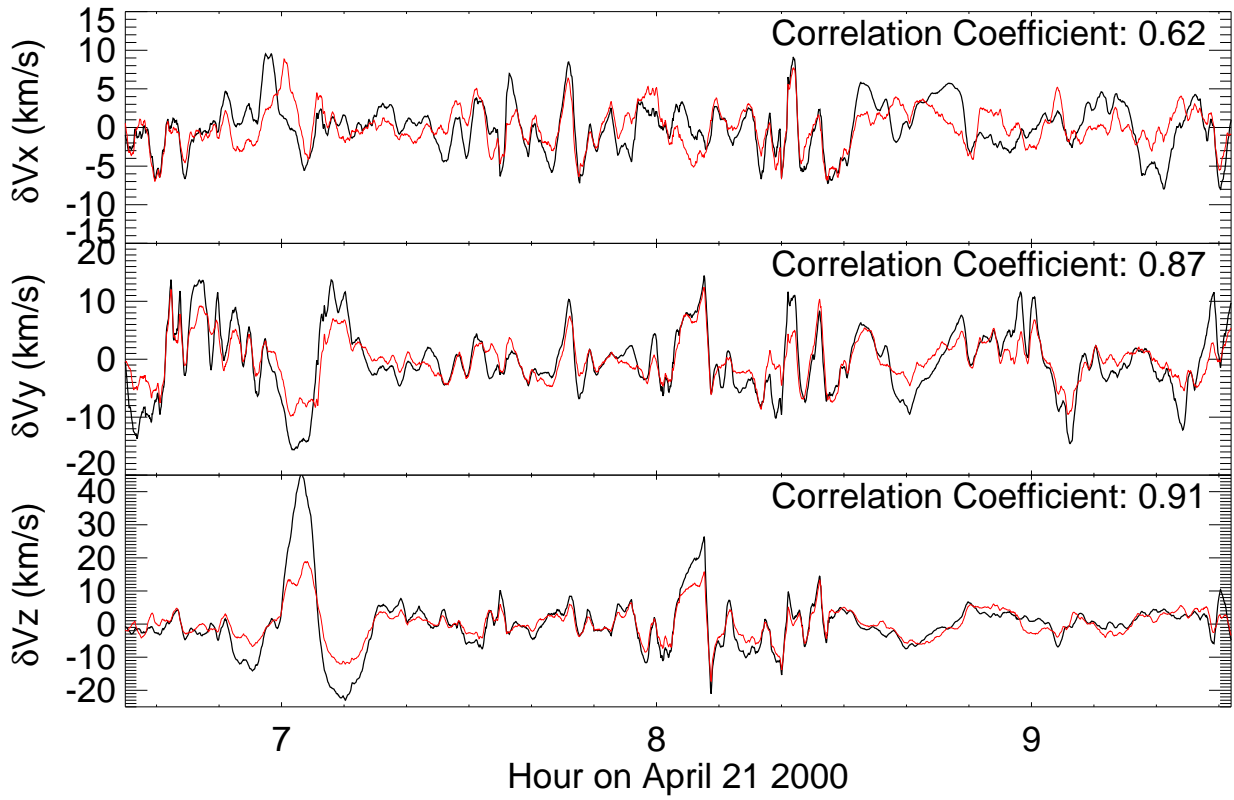}}
\caption{Evolution of the components of the Alfv\'{e}n velocity
(black) and proton velocity (red) perturbations for an Alfv\'{e}n
wave train observed by Wind on April 21 2000. The correlation
coefficients are also shown in each panel.} \label{fig.6}
\end{figure}

Figure~\ref{fig.6} shows the evolution of the components of the
Alfv\'{e}n velocity and proton velocity perturbations for an
identified Alfv\'{e}n wave train. The correlation coefficients for
the three components are 0.62, 0.87, and 0.91, respectively. These
high correlations, especially for the y and z-components, indicate
that Alfv\'{e}n waves are clearly present.

According to \cite{Cartwright2008}, structures with clear
Alfv\'{e}nic fluctuations should be rejected as possible flux ropes.
However, it might be possible that these Alfv\'{e}n waves were
generated inside the flux ropes. \cite{Okamoto2007} identified
transverse oscillations of fine-scale threads in solar prominences
and interpreted them as Alfv\'{e}n waves. If these waves are carried
by erupting prominences into the interplanetary space, they should
be observed in situ as correlated changes between the magnetic field
and velocity. In fact, signatures of Alfv\'{e}n waves have already
been found in magnetic clouds \citep{Marsch2009,Yao2010}. If
small-scale magnetic flux ropes are interplanetary manifestations of
small-scale solar eruptions \citep{Feng2007,Feng2008}, and
Alfv\'{e}n waves generated in small-scale solar activities
\citep{Mandrini2005,Innes2009} are carried outwards through their
eruptions to the interplanetary space, it will not be strange to
observed the Alfv\'{e}nic fluctuations in small-scale flux ropes. On
the other hand, if small-scale flux ropes are the products of local
magnetic reconnection in the solar wind
\citep{Moldwin2000,Cartwright2008}, Alfv\'{e}n waves could also form
since magnetic reconnection can generate the waves.

Although we do not exclude the possibility that some Alfv\'{e}n wave
trains might also be real magnetic flux ropes with intrinsic
Alfv\'{e}nic fluctuations, we decide to follow \cite{Cartwright2008}
and classify those Alfv\'{e}n wave trains identified by us into a
different group rather than flux ropes, since we can not provide any
evidence to prove that these waves are generated inside magnetic
flux ropes.

\subsection{Properties of magnetic reconnection at flux rope boundaries}

Among the 125 small-scale flux ropes in \cite{Feng2008}, we found 44
cases with clear Alfv\'{e}nic fluctuations and classified them into
Alfv\'{e}n wave trains. The rest 81 cases were classified into real
magnetic flux ropes. We found that 42\% (34/81) of the flux ropes
have signatures of magnetic reconnection at one or two of their
boundaries, while only 14\% (6/44) of the wave trains reveal
reconnection signatures at their boundaries. This result suggests
that the probability of magnetic reconnection is very high at flux
rope boundaries. The lower probability of reconnection associated
with wave trains than those associated with flux ropes also
indicates that the wave trains and flux ropes are likely to be
different structures, thus supporting our classification mentioned
above.

\begin{figure}
\centering {\includegraphics[width=0.8\textwidth]{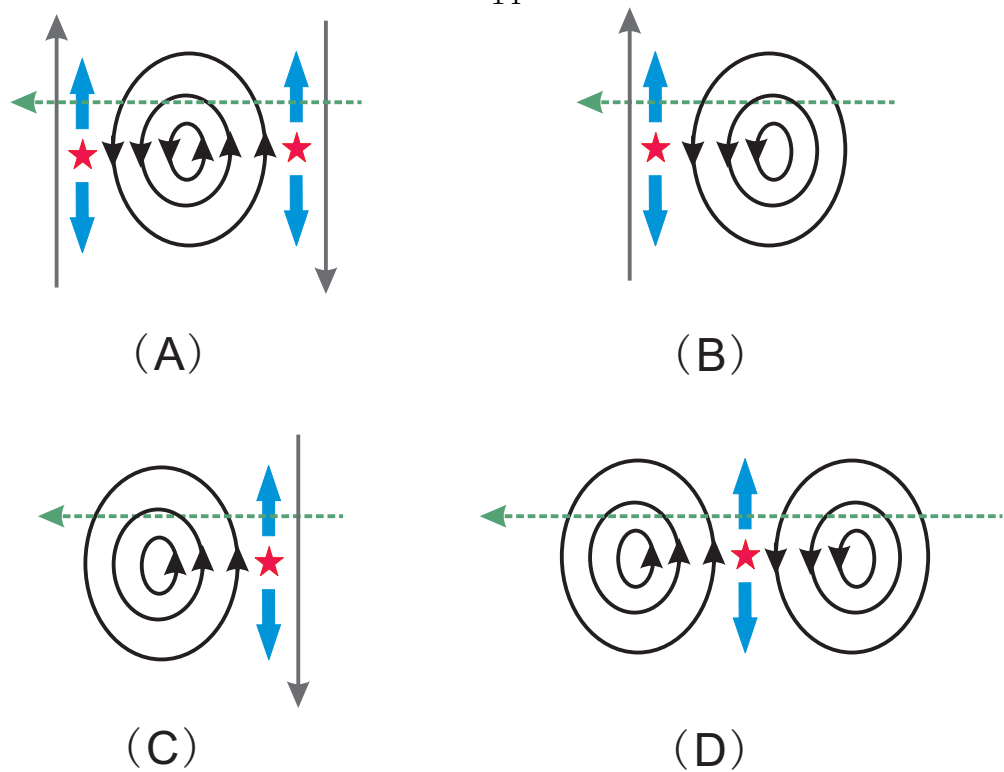}}
\caption{(A)-(C): Magnetic reconnection between interplanetary flux
ropes (black) and the ambient magnetic field (grey). (D): Magnetic
reconnection between two adjacent flux ropes. The stars and blue
arrows represent reconnecting X points and outflows, respectively.
The long dashed arrows indicate trajectories of \textit{Wind}. The
Sun is to the left. } \label{fig.7}
\end{figure}

In the highly turbulent solar wind, the ambient IMF is often
observed to change its direction away from the Parker spiral
locally. The well-organized magnetic field lines of a flux rope can
be expected to be obviously sheared with respect to the ambient IMF,
resulting in formation of a current sheet and sometimes reconnection
on the leading, trailing or both sides, as demonstrated in
Figures~\ref{fig.1}-\ref{fig.5} and sketched in
Figure~\ref{fig.7}(A)-(C). On the other hand, as will be discussed
in Section 3.4, the reconnection process might also be related to
the formation of flux ropes in the heliospheric current sheet.

\begin{figure}
\centering {\includegraphics[width=\textwidth]{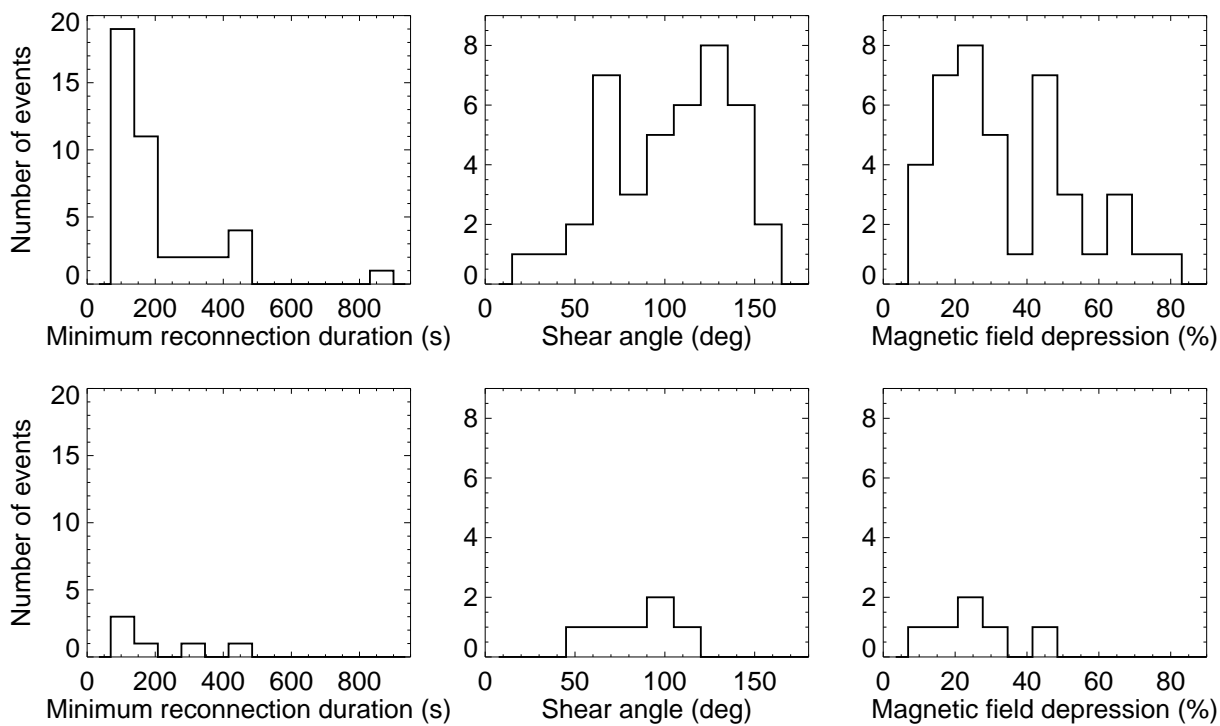}}
\caption{Distributions of the minimum reconnection duration (left),
shear angle of the magnetic field across the reconnection exhaust
(middle) and percentage of the magnetic field depression (right).
Results for the flux ropes and Alfv\'{e}n wave trains are presented
in the top and bottom panels, respectively. } \label{fig.8}
\end{figure}

The distributions of the minimum reconnection duration, shear angle
of the magnetic field across the reconnection exhaust and percentage
of the magnetic field depression for the reconnection events
associated with flux ropes and wave trains are presented in
Figure~\ref{fig.8}. The minimum reconnection duration, represented
by the observed exhaust duration, peaks at the first bin for both
the flux ropes and wave trains, indicating that in both cases most
of the reconnecting boundary layers are very thin.

We found that about 66\% of the reconnection events at flux rope
boundaries are associated with a magnetic field shear angle larger
than 90$^\circ$. Figure~\ref{fig.8} reveals clearly that large shear
angles are predominant for reconnection at flux rope boundaries. The
shear angle for reconnection associated with Alfv\'{e}n wave trains
ranges from 40$^\circ$ to 120$^\circ$. The average shear angles for
the two are 107$^\circ$ and 82$^\circ$, respectively. Our result
indicates that at boundaries of flux ropes the anti-parallel field
component usually exceeds the guide field component, resulting in
fast anti-parallel reconnection. Through a statistical study,
\cite{Gosling2007a} concluded that magnetic reconnection in the
solar wind occurs more frequently at shear angles smaller than
90$^\circ$ and is associated with relatively low reconnection rates.
Thus, considering the shear angles, magnetic reconnection in the
background solar wind is largely different from that at boundaries
of small-scale magnetic flux ropes.

\begin{figure}
\centering {\includegraphics[width=0.48\textwidth]{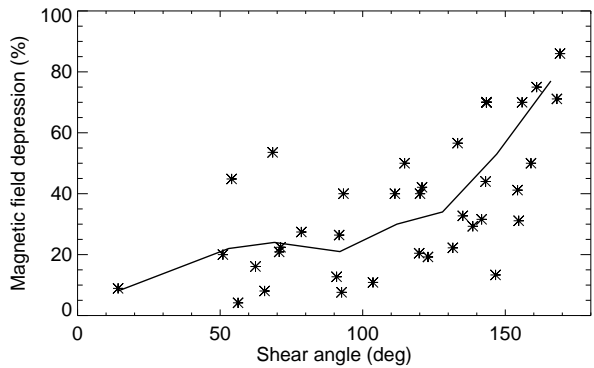}}
\caption{Relationship between the magnetic field depressions inside
reconnection exhausts and shear angles of the magnetic field across
the exhausts. Each asterisk represents one reconnection event at one
boundary of a flux rope. The line shows the average field
depressions in 9 shear-angle bins.} \label{fig.9}
\end{figure}

The depression of the magnetic field magnitude is larger than 20\%
for most reconnection exhausts associated with both flux ropes
(73\%) and wave trains (67\%). We also investigated the field
depression at flux rope boundaries without reconnection signatures,
and found that in most cases (64\%) the field depression is less
than 20\%. Figure~\ref{fig.9} shows the relationship between the
magnetic field depressions inside reconnection exhausts and shear
angles of the magnetic field across the exhausts. We divided the
shear angle into 9 bins, and averaged the field depression in each
bin (See the solid line in Figure~\ref{fig.9}). Although large
scatters are present, we can still see a clear trend of enhancement
in the magnetic field depression with increasing shear angle,
indicating that very strong magnetic field depressions (e.g., larger
than 35\%) are often associated with large shear angles and thus
anti-parallel reconnection.

\begin{figure}
\centering {\includegraphics[width=0.48\textwidth]{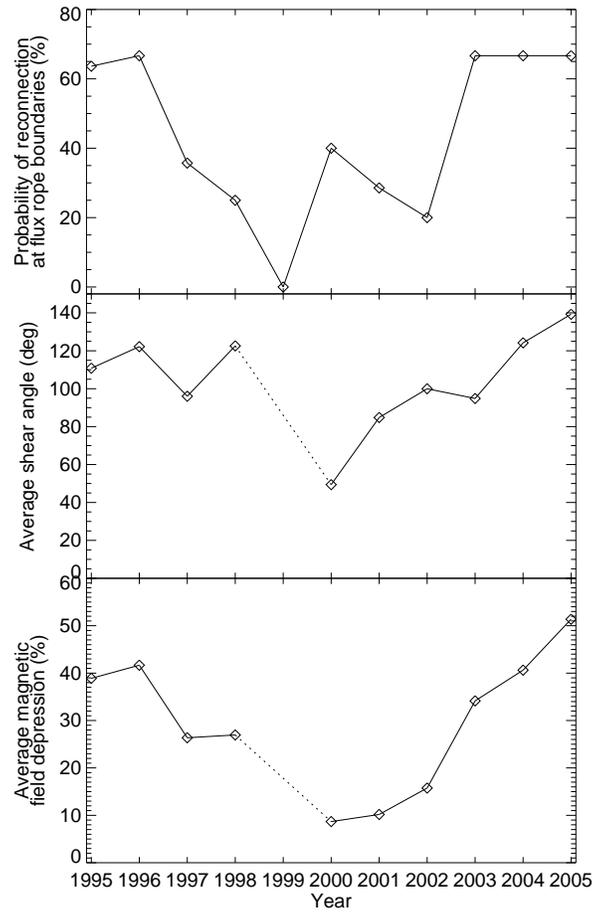}}
\caption{Probability of magnetic reconnection, average shear angle
across reconnection exhausts, and average magnetic field depression
inside reconnection regions at flux rope boundaries from 1995 to
2005. } \label{fig.10}
\end{figure}

We also investigated the occurrence rate of magnetic reconnection,
average shear angle across reconnection exhausts, and average
magnetic field depression inside reconnection regions at flux rope
boundaries through the year of 1995 to 2005. Figure~\ref{fig.10}
shows the result. We found that the probability of reconnection at
small-scale flux rope boundaries is relatively low around the solar
maximum and very high when approaching the solar minima. Note that
no reconnection event was found at boundaries of flux ropes in 1999.
A low occurrence rate was found in 1999 for both magnetic clouds and
small-scale flux ropes \citep{Cartwright2008,Feng2008}. The low
occurrence rate of magnetic clouds was suggested to be due to the
fact that the CME sources were migrated to higher latitudes of the
Sun in 1999 \citep{Gopalswamy2003}, and the low occurrence rate of
small-scale flux ropes might also be related to the solar activity
\citep{Feng2008}. Our result further indicates that at the year of
low occurrence of small-scale flux ropes, the occurrence rate of
reconnection at flux rope boundaries is also depressed.
Interestingly, the temporal variations of the average shear angle
and magnetic field depression, particularly the latter, also reveal
a similar trend. We speculate that at solar maximum the more complex
magnetic field structures both on the Sun and in the solar wind
might be responsible for these behaviors. More observational
analyses are needed to confirm and explain these results.

\subsection{Comparison between small-scale flux ropes and magnetic clouds}

So far we have demonstrated that signatures of magnetic reconnection
are present at boundaries of 42\% of the small-scale flux ropes.
\cite{Wei2003} found that a substantial parts of magnetic clouds
have boundary layers displaying a drop in the magnetic field
magnitude and a significant change of the field direction, as well
as properties of a high proton temperature, density and plasma beta.
They claimed that these signatures are manifestations of magnetic
reconnection through interactions between magnetic clouds and the
ambient medium. The small-scale flux rope boundaries studied by us
exhibit characteristics very similar to boundaries of large-scale
magnetic clouds, implying a similar interaction process between
these flux ropes and the magnetic field in the ambient medium. This
similarity might also suggest that some of the interplanetary
small-scale magnetic flux ropes have an origin similar to the
magnetic clouds \citep{Feng2007,Feng2008,Feng2009}.

However, \cite{Wei2003} did not check if the roughly Alfv\'{e}nic
plasma jets are prevalent at magnetic cloud boundaries. It would be
interesting to investigate if the Petschek-type reconnection
exhausts \citep[e.g.,][]{Gosling2005a} are present in the boundary
layers of magnetic clouds. \cite{Wei2003} mentioned that features of
magnetic reconnection regions are observed more frequently at the
leading boundaries of magnetic clouds than at the trailing
boundaries. In our Table~\ref{table1}, the numbers of reconnection
events occurring at the leading and trailing boundaries of
small-scale flux ropes are 20 and 21, respectively. So it seems that
the chance of reconnection is comparable at both boundaries of small
flux ropes.

Furthermore, we noticed that there are only 7 small-scale flux ropes
exhibiting signatures of reconnection at both boundaries. And more
than half of the small-scale magnetic flux ropes do not exhibit
signatures of magnetic reconnection at their boundaries. This result
might be related to magnetic field configurations at the interfaces
between flux ropes and the ambient medium, the restriction of the
temporal resolution, and the non-persistency of the reconnection
process. A small shear angle or a gentle variation of the magnetic
field across the interfaces may not be favorable for the occurrence
of magnetic reconnection. It is known that a higher time resolution
of the instrument leads to the discovery of more reconnection events
\citep{Gosling2008,Phan2009}, the 3-s resolution of the instruments
might not be sufficient to resolve some very narrow reconnection
exhausts present at flux rope boundaries. \cite{Wei2003} mentioned
that reconnection conditions might be weakened as the magnetic
reconnection at boundary layers of magnetic clouds proceeds,
resulting in the recovery of the frozen-in condition. The process
may continuously repeat and lead to intermittent magnetic
reconnection. This scenario may also be the case at the boundaries
of small-scale flux ropes, and could be one of the reasons of the
observed low occurrence rate of magnetic reconnection.

As mentioned in Section 2.2, Figure~\ref{fig.5} shows an example of
reconnection between two adjacent small-scale flux ropes. This
scenario is sketched in Figure~\ref{fig.7}(D). As the reconnection
proceeds, the two flux ropes could coalesce and thus grow in size,
as demonstrated in laboratory experiments and numerical simulations
\citep[e.g.,][]{Furno2005,Richard1989,Linton2006}. The two adjacent
small flux ropes might also be the small-scale counterpart of
multiple magnetic clouds
\citep[e.g.,][]{Wang2002,Wang2005,Xiong2007,Xiong2009}.

\subsection{Origin of interplanetary small-scale magnetic flux ropes}

There is a debate on the origin of interplanetary small-scale
magnetic flux ropes. \cite{Moldwin2000} and \cite{Cartwright2008}
suggested that these small flux ropes are produced through
interplanetary magnetic reconnection. While others proposed that
they are interplanetary manifestations of small-scale solar
eruptions \citep{Feng2007,Feng2008,Wu2008}. \cite{Wei2003} and
\cite{Pick2005} mentioned that the magnetic field lines of
relatively large flux ropes originating from the Sun could be peeled
off through magnetic reconnection away from the Sun. \cite{Feng2009}
reported a small flux rope followed by a reconnection exhaust, and
suggested that the small flux rope is produced through this peeling
off process.

Compared to \cite{Feng2009}, our statistic result provides many more
cases of small flux ropes with signatures of magnetic reconnection
at their boundaries. Our result reveals clearly that magnetic
reconnection is common at the interfaces between small flux ropes
and the ambient medium. However, it is still an open question
whether this reconnection is related to the formation of small flux
ropes or not. In the scenario suggested by \cite{Feng2009}, the flux
rope can be diminished in size due to reconnection with the ambient
magnetic field as it moves from the Sun to the Earth. However, such
shrinking has not been directly observed in interplanetary space. On
the contrary, large-scale flux ropes such as magnetic clouds usually
show a decrease in the measured plasma velocity as they pass the
spacecraft, indicating an expansion in size when moving away from
the Sun \citep[e.g.,][]{Lepping2006}. A recent study of
\cite{Cartwright2010} completed a comprehensive survey of
interplanetary small flux ropes observed between 0.3 and 5.5 AU
using the \textit{Helios}, \textit{IMP8}, \textit{WIND},
\textit{ACE}, and \textit{Ulysses} data, and found that on average
the size of small flux ropes expands rapidly within 1 AU and then
reaches equilibrium in the outer heliosphere. These results seem to
indicate that the expansion process dominates over the peeling off
process for flux ropes originating from the Sun, which is
inconsistent with the scenario that small flux ropes are produced
through the peeling off of magnetic field lines in the outer layers
of magnetic clouds.

In the Earth magnetospheric studies, multiple X line reconnection
\citep{Lee1985} is believed to be responsible for the observed flux
rope chains in the plasma sheet of the magnetotail
\citep[e.g.,][]{Slavin2003,Zong2004,Eastwood2005,Liu2009}. In
principle flux ropes could also be produced through a similar
process in interplanetary space. Observations reveal that
small-scale flux ropes lack a signature of expansion and are not
depressed in proton temperature, which are distinctly different from
magnetic clouds. Moreover, some small flux ropes were observed near
the sector crossing (Heliospheric current sheet crossing). These
observational facts seem to support the idea that the small flux
ropes are produced through magnetic reconnection across the
Heliospheric current sheet
\citep{Moldwin2000,Cartwright2008,Cartwright2010}. Using the similar
method of \cite{Cartwright2010}, we also investigated the time
difference between small-scale flux ropes listed in \cite{Feng2008}
and the nearest sector crossing. Among the 81 flux ropes, we could
only identify 35 events with clear sector crossing signature nearby.
While \cite{Cartwright2010} identified clear sector crossing
signature in 71 cases out of the total 91 flux ropes.
\cite{Cartwright2010} presented the distribution of the time to the
nearest sector crossing, and found a sharp peak with 17 flux ropes
observed within 6 hours of a sector crossing. Among the 35 events we
investigated, we found 9 flux ropes observed within 6 hours of a
sector crossing, but also found 18 flux ropes observed beyond one
day of the nearest sector crossing. The fact that some flux ropes
are very close to sector crossing and some others are far away from
clear sector crossing, as demonstrated both in \cite{Cartwright2010}
and our investigation, suggests that a subset of the small-scale
interplanetary flux ropes are likely to be produced through
reconnection across the heliospheric current sheet. If the current
sheet is locally tilted with respect to the passage of the
spacecraft, the typical reconnection signature of one plasma jet
within a current sheet should also be registered at boundaries of
newly formed flux ropes.

The scenario of solar origin could also be the case for some
interplanetary small-scale flux ropes. This idea, proposed by
\cite{Feng2007}, \cite{Feng2008}, and \cite{Wu2008}, was supported
by the continuous distribution of the duration of many small and
intermediate-scale flux ropes, the similar solar cycle dependence of
the occurrence rate between small flux ropes and magnetic clouds,
and the similar energy spectrum between small flux ropes and solar
flares. \cite{Wu2008} mentioned that due to the weaker magnetic
field inside small flux ropes than magnetic clouds, particles can
diffuse more easily and mix with each other between the flux ropes
and the ambient medium, leading to little change of proton
temperature across a small flux rope. The magnetic pressure inside a
small flux rope was proposed to be too small to drive expansion
\citep{Wu2008}. Moreover, the increase in the size of small-scale
flux ropes with increasing distance from the Sun discovered recently
by \cite{Cartwright2010} is similar to magnetic clouds, and thus
also supports the hypothesis of a similar origin of flux ropes with
different scales.

In conclusion, we think that interplanetary small-scale flux ropes
can be produced both in the solar wind and on the Sun. The
reconnection signatures we identified at flux rope boundaries could
either be related to the formation of flux ropes through
reconnection across the heliospheric current sheet, or result from
the subsequent interaction between flux ropes and the interplanetary
magnetic fields after the initial formation of the ropes.
Reconnection between magnetic clouds and interplanetary magnetic
fields can peel off some magnetic field lines but may not
efficiently reduce the size of magnetic clouds.

\section{Summary}

We have performed the first systematic study of the boundaries of
interplanetary small-scale magnetic flux ropes, and identified
signatures of magnetic reconnection at the boundaries. These
signatures generally include a drop in the magnetic field magnitude
and a plasma jet of $\sim$30~km~s$^{-1}$ in a field rotational
region. The reconnection regions often show additional properties
such as a local increase of the proton temperature, density and
plasma beta.

We have examined the magnetic field and plasma parameters of 125
small-scale flux ropes reported in \cite{Feng2008}, and found that
44 of them exhibit clear signatures of Alfv\'{e}nic fluctuations.
These flux ropes have been classified into Alfv\'{e}n wave trains
rather than real flux ropes. We have found that about 42\% of the
flux ropes and 14\% of the wave trains exhibit reconnection
signatures at one or two boundaries. About 2/3 of the reconnection
events at flux rope boundaries are associated with a shear angle
larger than 90$^\circ$, indicating fast anti-parallel reconnection.
Among the reconnection exhausts found at flux rope boundaries, about
73\% reveal a decrease by 20\% or more in the magnetic field
magnitude. More pronounced magnetic field depressions seem more
likely to be associated with larger shear angles. These results
suggest that reconnection at small-scale flux rope boundaries and in
the background solar wind is distinctly different, with the former
being much easier and faster.

We have also studied the occurrence rate of magnetic reconnection,
average shear angle across reconnection exhausts, and average
magnetic field depression inside reconnection regions at flux rope
boundaries through the year of 1995 to 2005, and found that all of
these parameters seem to be relatively low/small around solar
maximum and very high/large around solar minima.

The reconnection signatures we identified at small flux rope
boundaries could either be related to the formation of flux ropes
through reconnection across the heliospheric current sheet, or
result from the subsequent interaction between flux ropes and the
interplanetary magnetic fields after the ropes are ejected from the
Sun or formed in the solar wind. In the later case, our study
demonstrates that the boundaries of a substantial part of
interplanetary small-scale flux ropes resemble those of magnetic
clouds, and thus imply a similar interaction process between flux
ropes with different scales and the magnetic fields in the ambient
medium.

\begin{acknowledgements}
We thank the data providers, R. Lepping at NASA/GSFC, R. Lin at UC
Berkeley, and CDAWeb, for making the magnetic field and plasma data
(obtained by WIND/MFI and WIND/3DP) publicly available. We also
thank the anonymous referee for his/her effort to improve the paper.
H. Tian thanks Dr. H.-Q. Feng and A.-M. Tian for helpful
discussions. The authors are supported by the National Natural
Science Foundation of China under contracts 40874090, 40931055, and
40890162. The space physics group at PKU are also supported by the
Beijing Education Project XK100010404, the Fundamental Research
Funds for the Central Universities, and the National Basic Research
Program of China under grant G2006CB806305.
\end{acknowledgements}

\end{document}